\documentclass[a4paper,amsmath,amssymb]{jpconf}

\usepackage{graphicx}
\usepackage{hyperref}
\usepackage[sort&compress,square,comma,numbers]{natbib}
\usepackage{xcolor}
\usepackage{amsmath}
\usepackage{hyperref}

\begin{document}
\title{Accelerating the Inference of the Exa.TrkX Pipeline}

\author{Alina Lazar\textsuperscript{1}, Xiangyang Ju\textsuperscript{2}, Daniel Murnane\textsuperscript{2}, Paolo Calafiura\textsuperscript{2}, Steven Farrell\textsuperscript{2}, Yaoyuan Xu\textsuperscript{3}, Maria Spiropulu\textsuperscript{4}, Jean-Roch Vlimant\textsuperscript{4}, Giuseppe Cerati\textsuperscript{5}, Lindsey Gray\textsuperscript{5}, Thomas Klijnsma\textsuperscript{5}, Jim Kowalkowski\textsuperscript{5}, Markus Atkinson\textsuperscript{6}, Mark Neubauer\textsuperscript{6}, Gage DeZoort\textsuperscript{7}, Savannah Thais\textsuperscript{7}, Shih-Chieh Hsu\textsuperscript{8}, Adam Aurisano\textsuperscript{9}, Jeremy Hewes\textsuperscript{9}, Alexandra Ballow\textsuperscript{1}, Nirajan Acharya\textsuperscript{1}, Chun-yi Wang\textsuperscript{10}, Emma Liu\textsuperscript{11}, Alberto Lucas\textsuperscript{12}}

\address{
\textsuperscript{1}Youngstown State University,
\textsuperscript{2}Lawrence Berkeley National Lab, 
\textsuperscript{3}University of California-Berkeley,
\textsuperscript{4}California Institute of Technology, 
\textsuperscript{5}Fermi National Accelerator Laboratory,
\textsuperscript{6}University of Illinois Urbana-Champaign, 
\textsuperscript{7}Princeton University, 
\textsuperscript{8}University of Washington, 
\textsuperscript{9}University of Cincinnati,
\textsuperscript{10}National Tsing Hua University
\textsuperscript{11}University of California, Los Angeles
\textsuperscript{12}California State University, Monterey Bay
}

\ead{alazar@ysu.edu}

\begin{abstract}
Recently, graph neural networks (GNNs) have been successfully used for a variety of particle reconstruction problems in high energy physics, including particle tracking. The Exa.TrkX pipeline based on GNNs demonstrated promising performance in reconstructing particle tracks in dense environments. It includes five discrete steps: data encoding, graph building, edge filtering, GNN, and track labeling. All steps were written in Python and run on both GPUs and CPUs. In this work, we accelerate the Python implementation of the pipeline through customized and commercial GPU-enabled software libraries, and develop a C++ implementation for inferencing the pipeline. 
The implementation features an improved, CUDA-enabled fixed-radius nearest neighbor search for graph building and a weakly connected component graph algorithm for track labeling. GNNs and other trained deep learning models are converted to ONNX and inferenced via the ONNX Runtime C++ API.  The complete C++ implementation of the pipeline allows integration with existing tracking software. We report the memory usage and average event latency tracking performance of our implementation applied to the TrackML benchmark dataset. 
\end{abstract}

\section{Introduction}
Charged particle track reconstruction is an essential but computationally complex part of high energy physics (HEP) experiments. The ATLAS experiment uses a general-purpose detector at the  Large Hadron Collider (LHC), to collect data from proton-proton collisions in order to detect and investigate particles. The beam collision area is surrounded by layers of specialized detectors, each of the layers recording different properties of particles.

Charged particles resulted from the collisions travel away from the collision point through the magnetic field and ionize the material of the detectors, that in turn record 3D points left by the particles. The track reconstruction algorithm connects the recorded 3D points to reconstruct charged particle trajectories and extract the particle kinematics from reconstructed trajectories.

The computational complexity of the state-of-the-art tracking algorithms~\cite{ATLAS:2017kyn,CMS:2014pgm} grows faster than linear with beam intensity and detector occupancy. The future planned high-luminosity upgrade of the LHC will dramatically increase the rate of collisions in both the ATLAS and CMS experiments, delivering even more data. In this context, it becomes essential to develop fast tracking reconstruction algorithms to enable future scientific discoveries.

Possible approaches to speed up the computation of track reconstruction include: writing multithreading code that can be executed concurrently, using performant software technology such as optimized libraries, improving tracking strategy by adapting deep learning methods such as Graph Neural Networks (GNN), and integrating the software with modern hardware architectures, such as multi-core CPUs, GPUs and FPGAs.

In this work, we present an optimized GNN-based inference pipeline for track finding. The pipeline was implemented in Python. We optimize its computing performance through customized and commercial GPU-enabled software libraries and rewrite the whole pipeline in modern \textsc{C++} so that it can be integrated with existing experimental software framework.

\section{Description of the Exa.TrkX Pipeline}
The Exa.TrkX collaboration has developed an inference pipeline~\cite{Ju2021-rv} based on GNNs for track reconstruction. The pipeline, utilizing deep learning neural networks, includes multiple steps: preprocessing input data, learning embedding, building graphs, filtering edges, using a GNN to score graph edges, and connecting graph nodes according to edge scores in order to construct track candidates.

The preprocessing stage summarizes statistical features of the charge deposits associated with spacepoints and combines the cluster shape information with the positional information of spacepoints to form an input feature vector for each spacepoint that is suitable for machine learning. 

After the preprocessing stage, a multi-layer perceptron (MLP) embeds each spacepoint into a latent representation in which spacepoints coming from the same track are close by in Euclidean distance and away from others. To build the input graph needed for the GNN, it is necessary to construct a fixed-radius nearest neighbor (FRNN) graph in this embedded space. The FRNN graph has only edges connecting neighboring spacepoints residing within a constant radius (chosen as a hyperparameter). Once the FRNN graph is constructed, another MLP binary classifier is trained to select edges that are part of true tracks. The filtering stage is performed before training the more memory-intensive GNN. 

The next step consists of a GNN-based edge classification model that score all edges in the graph. The higher the edge score is, the more likely the edge connects two spacepoints that come from the same track. The GNN uses a message passing neural network framework, where node hidden states are concatenated to nearby edges and edge hidden states are aggregated and concatenated to connected nodes. The message passing algorithm is repeated eight times. To reconstruct track candidates, we select edges with scores above a threshold and use the weakly connected component graph algorithm to cluster spacepoints into groups; therefore, each group is a track candidate. Accuracy measurements, such as tracking efficiency and purity, are computed by matching true tracks with reconstructed tracks. 
The filtering and GNN stages require large memory resources, therefore, each stage of the pipeline is trained separately.

\section{Inference Accelerator Technologies on GPUs}
\label{sec:infopt}
The baseline Python-based inference pipeline uses the \textsc{PyTorch} 1.9.0 for embeddling and filtering and \textsc{TensorFlow} 2.5.0 for GNNs. Graph construction is implemented using the \textsc{RadiusGraph} transform from \textsc{PyG}~\cite{Fey2019-gs}. The resulting graph does not fully fit into the 16 GB GPU memory for the filtering stage, therefore, the inputs are divided in batches of a fixed size of 800,000 edges. The track labeling stage uses the \textsc{Scikit-learn} implementation of DBSCAN. 

We reported in~\cite{Ju2021-rv} that the baseline inference time depends almost linearly on the number of spacepoints in the event. However, building the graph (12 seconds) and labeling the tracks (over 2 seconds) are the two stages that take the most time. Starting with the baseline implementation, we employed a group of new accelerator technologies to optimize the graph construction and labeling stages. These accelerator technologies include the \textsc{Faiss}~\cite{Johnson2021-sa} and \textsc{FRNN}~\cite{hoetzlein2014fast} for nearest neighbor graph construction, and the CuGraph’s weakly connected components algorithm on the GPU~\cite{hricik2020using} for identifying track candidates, and the GPU-accelerated CuPy library~\cite{Nishino2017-kh} for data preprocessing and the PyTorch's mixed precision for inference speed-up.

\begin{figure}[h]
\begin{minipage}{17pc}
\includegraphics[width=1.0\textwidth]{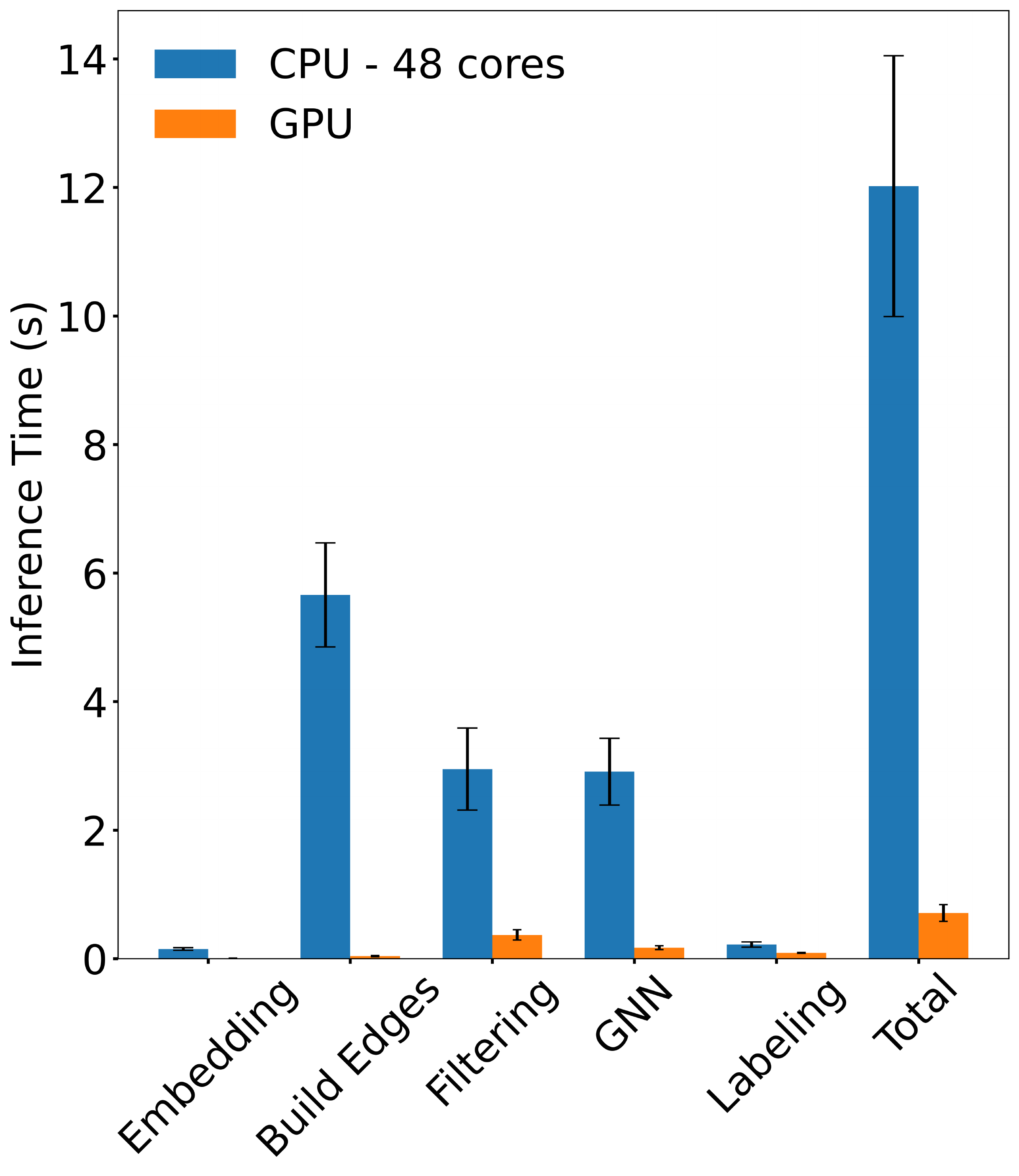}
\caption{Inference Time on GPU and CPU (48 cores). \label{fig:inftimegpu}}
\end{minipage}\hspace{2pc}
\begin{minipage}{17pc}
\includegraphics[width=1.0\textwidth]{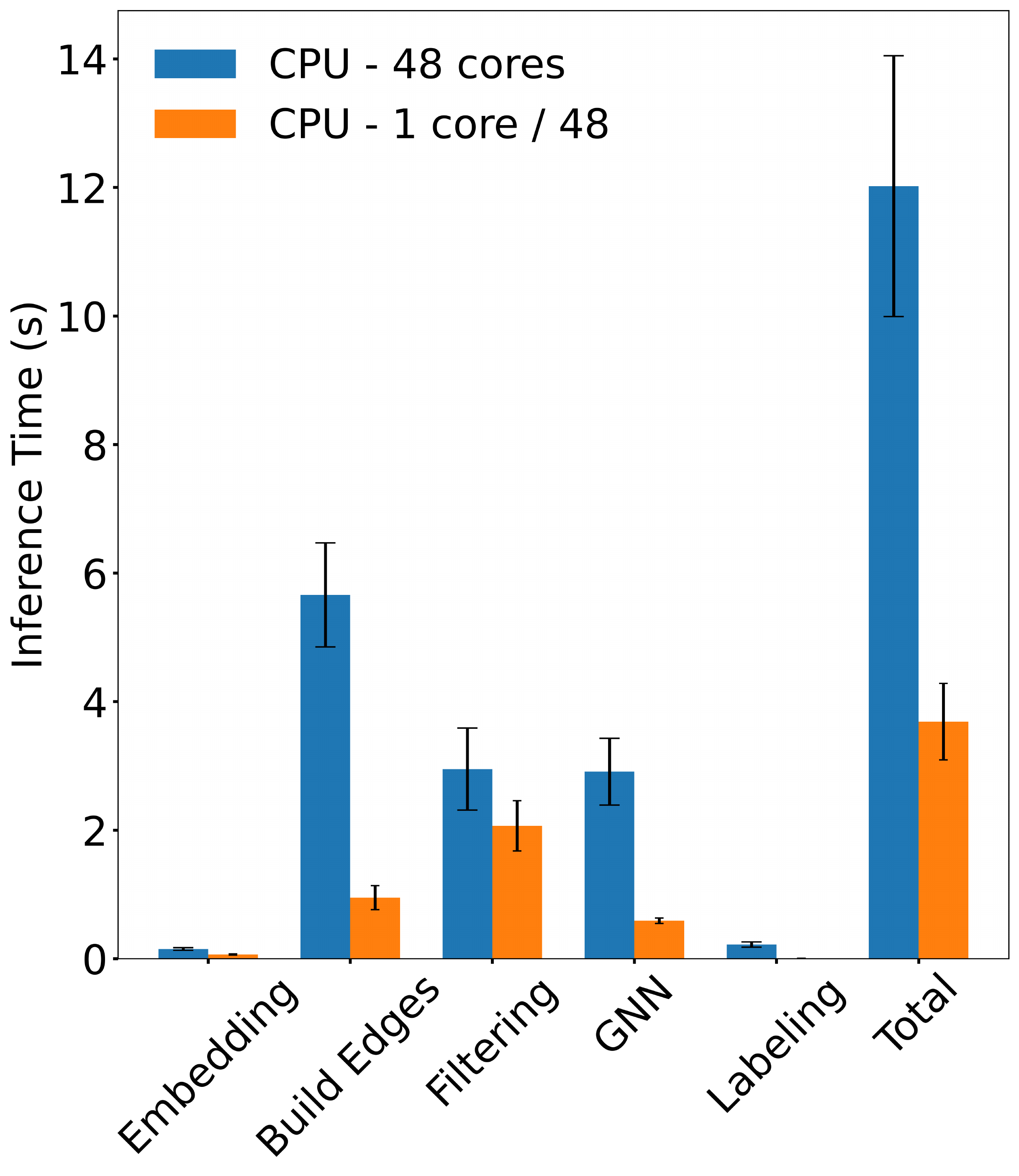}
\caption{Inference Time on CPU, one core vs 48 cores. \label{fig:inftimecpu}}
\end{minipage} 
\end{figure}

With those advanced accelerator technologies, we reduced the inference time from 15 seconds per event to $0.7$ seconds per event (as measured by the Python module time), on the GPU. 
We also evaluated the inference time on CPUs. The inference time is 179 seconds per event on a single CPU core and 12 seconds per event on one node with 48 cores. It is still faster to run the inference on Nvidia V100 GPUs compared with one full 48-cores CPU node for all stages, as shown in Figure~\ref{fig:inftimegpu}. The inference pipeline time does not scale linearly with the number of cores as shown in Figure~\ref{fig:inftimecpu}.

The following sub-sections discuss in detail the employment of accelerator technologies and their improvement. The experimental setup is described in Section~\ref{sec:setup}. 

\subsection{Graph Construction with Faiss}
Our baseline implementation employs the \textsc{RadiusGraph} method from the Pytorch Geometric library~\cite{Fey2019-gs} to build a graph from fixed-radius nearest neighbors (FRNN). Replacing this library with an implementation from the  Faiss library~\cite{Johnson2021-sa} cuts around 11 seconds from the average inference wall-time per event - from around 12 seconds to less than 1 second. Faiss has one the fastest similarity search implementations on both GPUs and CPUs. We use Faiss' similarity search to construct a brute-force kNN graph, by using a flat index and querying each spacepoint. A radius-based cut is then enforced on neighbor distances in the kNN neighborhoods. When datasets fit in the GPU memory, the Faiss kNN-search procedure is based on a well-parallelized exact kNN algorithm that takes advantage of GPU architecture. The optimized exhaustive search relies on the IndexFlat indexing structure that allows fast exact queries.

\subsection{Labeling for Track Identification}
On CPU, the Scikit-Learn DBSCAN algorithm \cite{Ester1996-xo,scikit-learn} is used to label all spacepoints based on connected edges passing some cut of the GNN. DBSCAN generates a kNN graph before computing and labeling clusters. However, this graph already exists, thus we provide a precomputed sparse adjacency matrix with distances given as one minus the sigmoid prediction of the GNN. Under the hood, DBSCAN with pre-computed adjacency uses a connected components algorithm. Sequential connected components labeling algorithms are typically performed with multiple passes over the graph nodes, adjusting each node's label according to its adjacency lists until each component or cluster has a single and unique label. A weakly connected components implementation for GPUs is available in the RAPIDS cuGraph library~\cite{hricik2020using}. This CUDA implementation~\cite{hawick2010parallel} incorporates data-parallel operations, several compute kernels, and uses the RAPIDS Memory Manager (RMM) library for memory allocation instead of the classical  \emph{cudaMalloc} and \emph{cudaFree}. The resulting code operates on the nodes and their adjacency-lists, is asynchronous and lock free, employs load balancing and speeds-up the inference pipeline with more than 2 seconds on average to 1.6 seconds. 

\subsection{Automatic Mixed Precision}
The PyTorch library provides mixed precision methods~\cite{micikevicius2017mixed} for training and inference, that works by matching some operations to half-precision floating point numbers and keeping other operations as single-precision for the float datatype. Instead of porting all operations to half precision, only the linear layers and convolutions are converted, while others, like tensor reductions (count, sum, average), that require the larger range of full-precision to sustain the model accuracy, are not. Mixed precision preserves the same accuracy as the full-precision model, but it reduces the model's runtime and the memory footprint. Mixed precision primarily benefits tensor core-enabled architectures, such as V100 and A100. This recipe provides (1.8-3x) speedup for our embedding and filtering MLP models.

\subsection{Fixed-Radius Nearest Neighbors}
The Faiss-based implementation of the graph construction improves timing over the PyTorch Geometric \texttt{RadiusGraph} method, however any exact kNN algorithm requires \textit{sorting} the list of neighbors of each query in order to find the nearest \emph{k} points. On the other hand, constructing a graph in a learned embedding space only requires an unsorted list of all the neighboring points residing within a certain distance from each query spacepoint. This situation is well-know as the fixed-radius nearest neighbors (FRNN) problem and can be solved using brute force, projections, cell techniques, and k-d trees. FRNN is applied in particle simulations of molecules or fluids with finite range interaction and to image segmentation of point clouds. The computational complexity of the brute force approach is quadratic $O(n^2)$ with respect to the number of spacepoints $n$. In comparison, sort-based approaches, such as projection and k-d trees, take $O(n\log{n})$. The biggest improvement is seen for grid-based techniques, which scale at worst-case as $O(n)$. However, accessing the cells scales exponentially with dimensionality $k$, as $O(3^k kn)$~\cite{BENTLEY1977209}. Nevertheless, in low dimensional spaces (2D and 3D), grid-based methods can be significantly faster than brute force. We implement a custom library\footnote{https://github.com/murnanedaniel/FRNN} that projects the relatively high-dimensional embedding space to three dimensions for FRNN search. We see a reduction of inference time down to 0.04 seconds compared to 0.54 seconds with Faiss and 12 seconds with the Pytorch Geometric implementation. 

\begin{table}[htb]
\centering
\vspace{-6mm}
\caption{Wall time of the Python-based Inference pipeline for the baseline and optimized implementations. The time is calculated with 500 events on an Nvidia Volta 100 GPU with a memory of 16 GB. The reported times are the average time and the standard deviation of the time in the unit of seconds. \label{table:timing_summary}}
\resizebox{\columnwidth}{!}{
\begin{tabular}{@{}l*{7}{l}}
\br
 &Baseline & Faiss & cuGraph & AMP & FRNN\\
\mr
Data Loading & $0.0022 \pm 0.0003$ & $0.0021 \pm 0.0003$ & $0.0023 \pm 0.0003$ & $0.0022 \pm 0.0003$ & $0.0022 \pm 0.0003$\\
Embedding & $0.02 \pm 0.003$ & $0.02 \pm 0.003$ & $0.02 \pm 0.003$ & $0.0067 \pm 0.0007$ & $0.0067 \pm 0.0007$\\
Build Edges & $12 \pm 2.64 $& $0.54 \pm 0.07$ & $0.53 \pm 0.07$ & $0.53 \pm 0.07$ & $0.04 \pm 0.01$\\
Filtering & $0.7 \pm 0.15$ & $0.7 \pm 0.15$ & $0.7 \pm 0.15$ & $0.37 \pm 0.08$ & $0.37 \pm 0.08$\\
GNN & $0.17 \pm 0.03$ &$0.17 \pm 0.03$ & $0.17 \pm 0.03$ & $0.17 \pm 0.03$ & $0.17 \pm 0.03$\\
Labeling & $2.2 \pm 0.3$ &$2.1 \pm 0.3$ & $0.11 \pm 0.01$ & $0.09 \pm 0.008$ & $0.09 \pm 0.008$\\
\mr
Total time & $15 \pm 3.$ & $3.6 \pm 0.6$ & $1.6 \pm 0.3$ & $1.2 \pm 0.2$ & $0.7 \pm 0.1$\\
\br
\end{tabular}
}
\end{table}

\subsection{Experiments \label{sec:setup}}
The hardware platforms, from the Ohio Supercomputer Center~\cite{OhioSupercomputerCenter1987}, used to run the inference experiments and measure performance have the following configuration: Nvidia Volta GPUs with 16 GB on-board memory, and Dual Intel Xeon 8268s (Cascade Lakes) CPUs with 48 cores per node and 2.9 GHz clock frequency and 4 GB memory per core. \textsc{CUDA} toolkit v11.1, CUDA v11.1 and the corresponding developer drivers 450.80 for the Nvidia GPUs are installed on this system. The peak memory consumption of the Exa.TrkX pipeline is about 15.7 GB on GPUs and about 11 GB on the CPUs. The results for the optimized GPU implementation (last column in Table~\ref{table:timing_summary}) show that the filtering and GNN steps are the biggest targets for further optimizations with the goal to surpass traditional algorithms in terms of inference time~\cite{Collaboration2019-rx}. GNNs are still considered new methods and optimization techniques are being actively developed.

\section{C++ implementations Using ONNX Runtime}
We convert the Python pipeline implementation to a C++ implementation\footnote{https://github.com/exatrkx/exatrkx-acat2021\label{github}} that runs in multithreading environments and can be easily integrated into C++-based tracking reconstruction frameworks. 

ONNX, Open neural network exchange~\cite{bai2019}, is a standard open format to represent machine and deep learning models. It was created as a platform of interoperability between deep learning frameworks. ONNX Runtime, the inference engine built based on the ONNX standards, is highly optimized for low-latency inference and it supports multiple backends and optimization methods. 

ONNX has built-in support for some of most of the popular deep learning layers and operators, including multilayer perceptrons and convolutational networks. However, the implementation of tensor reduction operators, such as the \textsc{scatter\_add} operator, has issues and are not optimized for either CPUs or GPUs as of writing. As a result, the physics performance of GNN's ONNX model is compromised. 

In addition to the deep learning models, we ported the Python implementation of FRNN (for graph construction) to CUDA and \textsc{libtorch}. Track labeling uses \textsc{libCuGraph} (the C++ API for cuGraph) from the RAPIDS AI~\footnote{https://rapids.ai/about.html}. One of the C++ conversion challenges was to construct an environment compatible with all libraries (\textsc{libtorch}, \textsc{PyG}, \textsc{ONNX Runtime}, and \textsc{RAPIDS AI})\footnote{From January 2022, compatibility is greatly simplified by RAPIDS AI using CUDA Enhanced Compatibility, which only requires \textit{minimum} CUDA runtime versions, rather than \textit{exact} version matching}. To solve this problem we built a Docker container with all the dependencies. The Dockerfile is available in the Exa.TrkX github repository.

\section{Integration of Exa.TrkX Inference with ACTS}
ACTS~\cite{Gessinger:2020nne} is a generic framework for tracking reconstruction implemented with modern software concepts and inherently designed for parallel architectures. Its core tracking finding algorithms are based on the Kalman filter method. We integrate our C++ implementation of the pipeline into ACTS so as to compare our ML-based track finding algorithm with conventional algorithms. To that end, we implemented our pipeline as a standalone plugin component in the ACTS and defined a new generic interface for track finding algorithms. This interface allows users to implement any track finding algorithm as long as the algorithm takes a vector of spacepoints as inputs and produces a list of track candidates, such as the Exa.TrkX pipeline. During the integration, we created an exemplar tracking reconstruction chain in that our Exa.TrkX replaces the Kalman filter method to find track candidates.

\section{Conclusions and Future Work}
We optimized the Exa.TrkX inference pipeline with state-of-the-art inference accelerator technologies and observed ~15x speedup. The Python GPU-based inference runs in sub-second average time for each event. Running inference on multiple CPU cores speeds up running the pipeline, but it is still ~17x slower than on the GPU, but note that there are still many more opportunities for optimization on CPUs. We have converted the inference pipeline to run in C++, in a multithreading environment. This allows the integration of the pipeline with other tracking frameworks such as ACTS.

In future work, one goal is to optimize the performance of the ONNX Runtime, by using half-precision ONNX models in the C++ inference pipeline. Choosing TensorRT as the provider for the ONNX Runtime requires special compilation of the ONNX Runtime but may provide further speedups. We are working with NVIDIA developers and other HEP stakeholders to develop new methods for accelerating the networks and operators required for large GNNs, as these are the main targets for further optimization in order to surpass traditional tracking algorithm inference time. Other goals are to reduce the number of software library dependencies for Python and C++ and to integrate the pipeline with the NVIDIA Triton Inference Server for further scalable optimizations.

\ack
This research was supported in part by the U.S. Department of Energy’s Office of Science, Office of High Energy Physics, of the US Department of Energy under Contracts No. DE-AC02-05CH11231 (CompHEP Exa.TrkX) and No. DE-AC02-07CH11359 (FNAL LDRD 2019.017); and by the National Science Foundation under Cooperative Agreement OAC-1836650. This research was supported in part by the Exascale Computing Project (17-SC-20-SC), a joint project of the Office of Science and National Nuclear Security Administration. Nirajan Acharya, Emma Liu and Alberto Lucas were supported by the XSEDE EMPOWER program under National Science Foundation grant number ACI-1548562.

This research used resources of the National Energy Research Scientific Computing Center (NERSC), a U.S. Department of Energy Office of Science User Facility located at Lawrence Berkeley National Laboratory, operated under Contract No. DE-AC02-05CH11231 and the Ohio Supercomputer Center (OSC).
 
\bibliographystyle{iopart-num}
\bibliography{IOPConf}

\end{document}